\documentclass[conference]{IEEEtran}
\IEEEoverridecommandlockouts
\usepackage{xcolor}

\newcommand{\comment}[1]{}
\usepackage{float}
\newcommand{\up}{\ensuremath{^{\uparrow}}}
\newcommand{\down}{\ensuremath{^{\downarrow}}}

\usepackage{array}
\usepackage{multirow}
\usepackage{threeparttable}
\usepackage{pifont}
\usepackage{cite}
\usepackage{amsmath,amssymb,amsfonts}
\usepackage{algorithmic}
\usepackage{graphicx}
\usepackage{subcaption}
\usepackage{textcomp}
\usepackage{hyperref}
\usepackage{booktabs}
\usepackage{amsmath}
\usepackage{orcidlink}
\usepackage[normalem]{ulem}

\newcommand{\cmark}{\ding{51}}
\newcommand{\xmark}{\ding{55}}

\newcommand{\best}[1]{\mathbf{#1}}
\newcommand{\second}[1]{\underline{#1}}

\newcommand{\LR}{\textbf{L} & \textbf{R}}

\def\BibTeX{{\rm B\kern-.05em{\sc i\kern-.025em b}\kern-.08em
    T\kern-.1667em\lower.7ex\hbox{E}\kern-.125emX}}

\begin{document}
\title{
It Takes Few %\martyna{Bits} 
to TANGO: A Quantized Distributed Model for Binaural Speech Enhancement \\
% {\footnotesize \textsuperscript{*}Note: Sub-titles are not captured for https://ieeexplore.ieee.org  and

}

\author{
\IEEEauthorblockN{
\begin{tabular}{c}
Zahra Benslimane\,\orcidlink{0009-0005-9846-9767}\IEEEauthorrefmark{1}\IEEEauthorrefmark{2},
Pierre Chouteau\IEEEauthorrefmark{1},
Martyna Poreba\,\orcidlink{0000-0002-5102-7735}\IEEEauthorrefmark{1},\\
Fabrice Auzanneau\,\orcidlink{0000-0002-4441-9239}\IEEEauthorrefmark{1},
Michal Szczepanski\,\orcidlink{0009-0000-9061-4396}\IEEEauthorrefmark{1},
Fabian Chersi\,\orcidlink{0000-0003-1257-2607}\IEEEauthorrefmark{1},
Romain Serizel\,\orcidlink{0000-0002-6848-0114}\IEEEauthorrefmark{2}
\end{tabular}
}

\vspace{0.4em}

\IEEEauthorblockA{
\IEEEauthorrefmark{1}
Universit\'e Paris-Saclay, CEA, List, F-91120 Palaiseau, France
}

\IEEEauthorblockA{
\IEEEauthorrefmark{2}
Universit\'e de Lorraine, CNRS, Inria, LORIA, F-54000 Nancy, France
}

\IEEEauthorblockA{
zahra-hafida.benslimane@cea.fr
}
}

%%%%%%%%%%%%%%%%%%%%%%%%%%%%%%%%%%%%%%%%%%%%%%%%%%%%%%%%%%%%%%%%%

\comment{
% Anonymous review placeholder - remove upon acceptance
\author{
  \IEEEauthorblockN{Anonymous Authors}
  \IEEEauthorblockA{
    \textit{Anonymous Institution} \\
    City, Country \\
    anonymous@institution.example
  }
}
}
%%%%%%%%%%%%%%%%%%%%%%%%%%%%%%%%%%%%%%%%%%%%%%%%%%%%%%%%%%%%%%%

\maketitle

\begin{abstract}
%%%%%%%%%%%%%%%%%%%%%%%%%%%%%%%%%%%%%%%%%%%%%%%%%%%%%%%%%%%%%%%%%%%%%%%%%%%%%%%%%%%%%%%%
%Martyna
%\textcolor{red}{Martyna: Remarque generale: Contribution scientifique pas assez saillante, Répétitions lexicales (computentional cost, memory cost ect.), manque à la fin un resultat chiffré. Abstract donne impression: We tested quantization, then simplifed architecture, then compressed more. En effet, il y a au moins 3 contributions à mettre en valeur.}\zahra{true i agree..}\\
%%%%%%%%%%%%%%%%%%%%%%%%%%%%%%%%%%%%%%%%%%%%%%%%%%%%%%%%%%%%%%%%%%%%%%%%%%%%%%%%%%%%%%%%
Neural network-based multichannel speech enhancement systems achieve strong enhancement performance, but their computational and memory requirements limit deployment on resource-constrained devices. This paper investigates low-precision inference for TANGO, a hybrid distributed binaural speech enhancement system combining neural mask estimation with spatial filtering. We evaluate post-training quantization and quantization-aware training for the neural components, and analyze how quantization errors in the mask estimators propagate through the downstream spatial filtering stage. Our analysis shows that, although quantization degrades intermediate mask estimates, the spatial filtering stage compensates for most quantization-induced errors. Leveraging this robustness, we simplify TANGO into MN-TANGO, reducing both model size and computational complexity while maintaining comparable final performance. By combining INT8 weight-and-activation quantization with ERB compression and grouped recurrent layers, the most compact MN-TANGO reaches 4.65 MMAC/s and 0.177 MB.

%We further identify a simplified TANGO configuration that improves the balance between computational complexity and enhancement performance, and combine low-precision inference with grouped recurrent layers and ERB feature compression to obtain a final architecture with reduced memory footprint and lower computational cost. 

\end{abstract}

\begin{IEEEkeywords}
Speech enhancement, quantization-aware training, recurrent neural networks, low-compute.
\end{IEEEkeywords}

\section{Introduction}
Deep learning approaches to speech enhancement (SE) have achieved strong performance, but they often rely on large and computationally expensive models. This limits their deployment on resource-constrained devices, such as embedded systems and hearing aids, where low-latency and low-power inference are critical. To address this limitation, a growing body of work has investigated model compression for neural SE.

Early studies mainly focused on reducing model storage through weight compression. Wu et al.~\cite{Wu_2019} combined channel pruning with k-means clustering to quantize the weights of a time-domain fully convolutional network. Similarly, Tan and Wang~\cite{9437977} applied sparse regularization, iterative pruning, and k-means-based quantization to several architectures, including temporal convolutional networks, and gated convolutional recurrent networks (GCRNs). Other works explored reduced floating-point representations. Hsu et al.~\cite{Hsu2018ASO} introduced the Exponent-Only Floating-Point Quantized Neural Network (EOFP-QNN), which quantizes the mantissa and exponent separately. Lin et al.~\cite{9648024} went further by discarding the mantissa entirely and retaining only the sign and exponent bits, achieving about $81\%$ model compression.

While these methods reduce memory footprint, they primarily target weights. Activations, inputs, and outputs often remain in floating point, so inference may still require costly floating-point arithmetic. This limits efficiency on low-power hardware, such as microcontrollers and neural processing units, which are typically optimized for integer pipelines such as INT8. To address this limitation, Fedorov et al.~\cite{fedorov20_interspeech} proposed TinyLSTMs, combining structured pruning with quantization-aware training (QAT)~\cite{jacob2017quantizationtrainingneuralnetworks} to quantize both weights and activations to 8 bits, while keeping the model outputs on 16-bit. Recent studies further showed that activation and I/O standard QAT can be more challenging than weight quantization alone, especially at high input signal-to-noise ratios (SNRs). To mitigate this effect,~\cite{cohen23_interspeech} introduced a residual correction branch to compensate for quantization errors. This approach was later extended to source separation using a knowledge-distillation-based loss for quantization-sensitive samples~\cite{10591369}.

% \begin{figure*}[ht]
%     \centering
%     \includegraphics[width=\textwidth]{media/tangos2.pdf}
%     \caption{Overview of the evaluated TANGO variants: (a) original two-stage TANGO with SN-DNN followed by MN-DNN processing, (b) MN-TANGO with only the MN-DNN stage, and (c) inverted TANGO with MN-DNN processing before the SN-DNN stage using either $C^\dagger$ or $C^\star$ as second-stage input.}
%     \label{fig:tango_variants}
% \end{figure*}

\begin{figure*}[ht]
    \centering
    \begin{subfigure}[b]{0.32\textwidth}
        \centering
        \includegraphics[width=\textwidth]{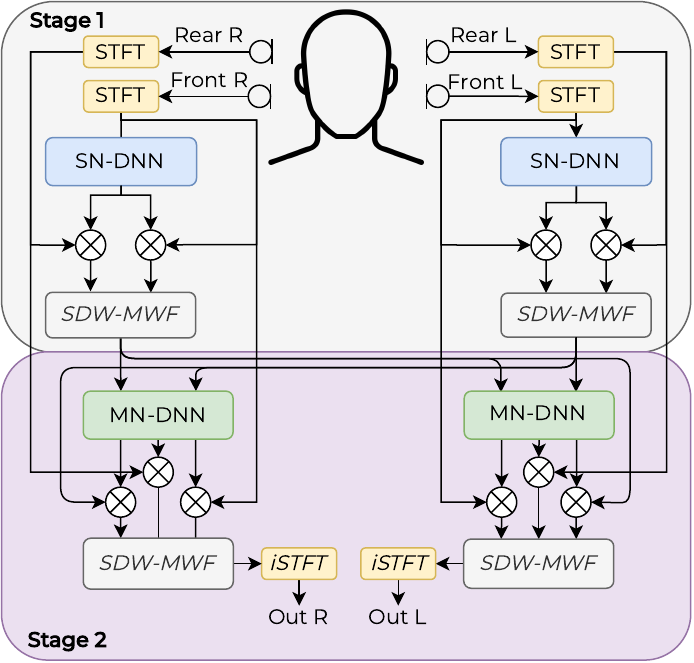}
        \caption{Original two-stage TANGO}
        \label{fig:tango}
    \end{subfigure}%
    \hfill
    \begin{subfigure}[b]{0.32\textwidth}
        \centering
        \includegraphics[width=\textwidth]{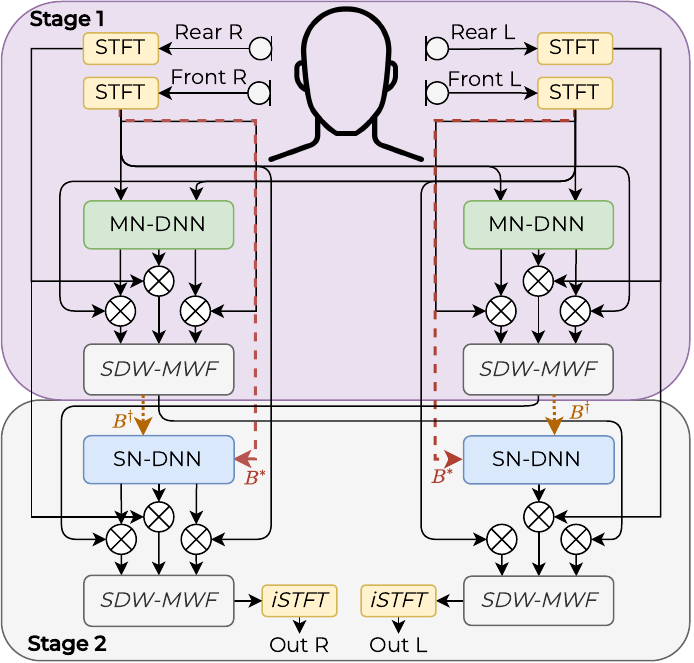}
        \caption{Inverted TANGO}
        \label{fig:inv_tango}
    \end{subfigure}%
    \hfill
    \begin{subfigure}[b]{0.32\textwidth}
        \centering
        \includegraphics[width=\textwidth]{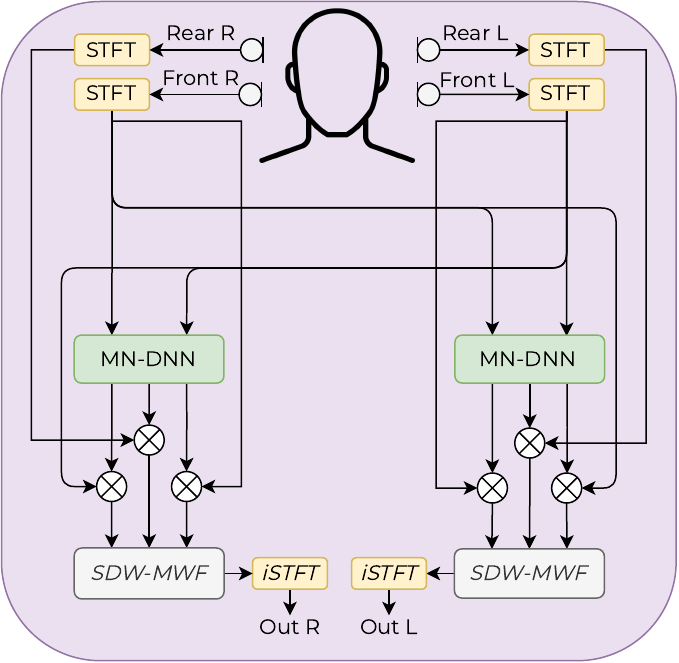}
        \caption{MN-TANGO}
        \label{fig:mn_tango}
    \end{subfigure}
    \caption{Overview of the evaluated TANGO variants: (a) original two-stage TANGO with SN-DNN followed by MN-DNN processing, (b) inverted TANGO with MN-DNN processing before the SN-DNN stage, and (c) MN-TANGO with only the MN-DNN stage. In (b), the dotted lines indicate the two alternative inputs to the second-stage SN-DNN: ($B^\dagger$) uses the output of the first spatial filtering stage, whereas ($B^\star$) uses the local reference signal.} 
    \label{fig:tango_variants}
\end{figure*}

Despite these advances, most quantization studies for SE focus on single-channel, purely neural models, leaving hybrid multichannel systems largely unexplored. In this work, we show that the hybrid neural-spatial structure of TANGO~\cite{9466439} makes it robust to low-precision neural inference. We focus on quantizing the neural network rather than the spatial filter, since the neural component accounts for most of the computational cost~\cite{Benslimane25}. Although quantization degrades intermediate mask estimates, the downstream spatial filtering stage compensates for most quantization-induced errors. Our results highlight three main findings: (i) the final spatial filtering stage provides most of the enhancement gains, (ii) spatial filtering mitigates most of the degradation introduced by quantization, and (iii) the original two-stage architecture can be simplified into its second stage (MN-TANGO) while maintaining comparable final performance. Building on these findings, we combine MN-TANGO with QAT, Equivalent Rectangular Bandwidth (ERB) compression, and grouped recurrent layers to obtain a compact low-compute model for distributed binaural SE.

\section{Background}
%This section investigates the sensitivity of the original TANGO model to low-precision inference. \zahra{à travailler}

%\subsection{Baseline model}
\subsection{Baseline TANGO Architecture}
\label{ssec:baseline_model}

%TANGO is used as the baseline system~\cite{9466439}.
TANGO~\cite{9466439} is used as the baseline architecture in this study. Like other hybrid SE systems, TANGO combines the representation learning capabilities of deep neural networks with the spatial filtering properties of classical beamformers. This hybrid design makes it particularly relevant for studying the impact of quantization on both neural and signal-processing components. In the first stage, each ear-node independently estimates speech and noise time-frequency masks using a single-node DNN (SN-DNN). These masks are used to estimate speech and noise spatial covariance matrices, from which a GEVD-based~\cite{6730918} speech-distortion-weighted multichannel Wiener filter (SDW-MWF)~\cite{doclo2007frequency} is derived and applied as a spatial filter.
The resulting ear-specific compressed signal is then transmitted to the contralateral ear-node. In the second stage, a multi-node DNN (MN-DNN) refines the mask estimates by exploiting both local signals and the exchanged representation. A final SDW-MWF then generates the enhanced binaural output. The overall architecture is illustrated in Fig.~\ref{fig:tango_variants}\subref{fig:tango}.

\begin{table*}[t]
\centering

\caption{Model comparison with quantization scheme, precision configuration, memory, and left/right ear scores. W, A, and I/O denote weights, activations, and input/output tensors, respectively.}
\label{tab:quant_lr_metrics}

\begin{tabular}{
    lcccc
    *{5}{cc}
}
\toprule

\multirow{2}{*}{\textbf{Quant. scheme}}
& \multicolumn{3}{c}{\textbf{Precision}}
& \textbf{Memory}\down
& \multicolumn{2}{c}{\textbf{SI-SIR}\up}
& \multicolumn{2}{c}{\textbf{SI-SDR}\up}
& \multicolumn{2}{c}{\textbf{SI-SAR}\up}
& \multicolumn{2}{c}{\textbf{STOI}\up}
& \multicolumn{2}{c}{\textbf{PESQ}\up} \\\cmidrule(lr){2-4}\cmidrule(lr){5-5}\cmidrule(lr){6-7}\cmidrule(lr){8-9}
\cmidrule(lr){10-11}\cmidrule(lr){12-13}\cmidrule(lr){14-15}

& \textbf{W} & \textbf{A} & \textbf{I/O}
& \textbf{MB}
& \LR
& \LR
& \LR
& \LR
& \LR \\
\midrule

Noisy
& $-$ & $-$ & $-$
& $-$
& $0.0$ & $-4.0$
& $-0.6$ & $-4.6$
& $-$ & $-$
& $0.68$ & $0.56$
& $1.14$ & $1.10$ \\ 
\midrule
Float32
& FP32 & FP32 & FP32
& $4.03$
& $22.8$ & $26.2$
& $4.7$ & $5.0$
& $5.0$ & $5.1$
& $0.842$ & $0.850$
& $1.731$ & $1.770$ \\

\midrule

DPTQ
& INT8 & FP32 & FP32
& $1.01$
& $18.4$ & $20.9$
& $2.7$ & $2.9$
& $3.6$ & $3.6$
& $0.811$ & $0.813$
& $1.585$ & $1.614$ \\ 

\midrule

QAT
& INT8 & FP32 & FP32
& $1.083$
& $22.8$ & $26.2$
& $4.7$ & $5.0$
& $5.0$ & $5.1$
& $0.843$ & $0.851$
& $1.729$ & $1.765$ \\

QAT
& INT8 & INT8 & INT16
& $1.083$
& $23.0$ & $25.9$
& $3.7$ & $4.5$
& $4.0$ & $4.6$
& $0.828$ & $0.842$
& $1.735$ & $1.753$ \\
\bottomrule

\end{tabular}
\end{table*}

\subsection{Neural Network Quantization}
\label{ssec:quantization}

%%%%%%%%%%%%%%%%%%%%%%%%%%%%%%%%%%%%%%%%%%%%%%%%%%%%%%%%%%%%%%%%%%%%%%%%%%%%%%%%
% sinon simplefied version????
%In this work, quantization is used to reduce the precision of the neural mask estimators while keeping the signal-processing operations in floating point. We consider two quantization strategies. Dynamic post-training quantization (DPTQ) is applied after floating-point training: the weights are stored in low precision, while activation ranges are determined dynamically at inference time. This approach is simple and does not require retraining, but it does not allow the model to adapt to quantization-induced errors.

%Quantization-aware training (QAT) instead simulates low-precision inference during training by inserting fake-quantization operators in the forward pass. These operators emulate rounding and clipping effects while keeping the computation differentiable through a straight-through estimator. As a result, the model can adapt its parameters to the numerical constraints that will be encountered at inference time. 
%%%%%%%%%%%%%%%%%%%%%%%%%%%%%%%%%%%%%%%%%%%%%%%%%%%%%%%%%%%%%%%%%%%%%%%%%%%%%%ù

Quantization maps floating-point values to a lower-precision integer representation. This reduces memory requirements and can improve inference efficiency, particularly on hardware that supports low-precision arithmetic. In neural networks, it can be applied both to the weights and to the intermediate activations.

In uniform affine quantization, values are mapped from a selected floating-point range to a finite set of integer levels determined by the target bit-width. The scale controls the spacing between these levels, while clipping ensures that values outside the selected range remain representable. In dynamic post-training quantization (DPTQ), quantization is applied after floating-point training: the layer weights are stored in low precision, while activations are quantized dynamically at runtime according to their observed range. In contrast, QAT simulates low-precision inference during training by inserting fake-quantization modules in the forward pass. Since rounding is non-differentiable, gradients are approximated using a straight-through estimator, allowing the model to adapt to quantization errors before deployment~\cite{jacob2017quantizationtrainingneuralnetworks}.

\label{quantized_tango}

\section{Toward Low-Compute Quantized Tango}
\subsection{From TANGO to MN-TANGO}
\label{ssec:reduced_e2e}

To better understand the contribution of each TANGO stage, we investigate alternative architectures that either reorganize or simplify its processing stages, as shown in Fig.~\ref{fig:tango_variants}. We first evaluate two inverted configurations, denoted as ($B^\star$) and ($B^\dagger$) and illustrated in Fig.~\ref{fig:tango_variants}(b). Both variants start with a cross-node exchange of the reference signals and apply an MN-DNN before the first spatial filtering stage. They differ in the input used by the second-stage SN-DNN: ($B^\dagger$) uses the output of the first spatial filtering stage, whereas ($B^\star$) uses only the local reference signal of the corresponding node. These variants allow us to assess whether performing multi-node processing earlier in the pipeline improves enhancement performance. We further investigate a simplified MN-only configuration, shown in Fig.~\ref{fig:tango_variants}(c), that removes the SN-DNN stage entirely. This variant tests whether the first single-node stage is necessary once inter-node information is available, or whether most of the final enhancement can be preserved by combining multi-node mask estimation with the final spatial filtering stage. In the following, we refer to this configuration as MN-TANGO.

%downstream filter is able to exploit its spatial structure of the binaural signals and compensate for imperfect mask estimates

%To reduce the computational complexity of TANGO while better understanding the contribution of each processing stage, we investigate alternative architectures that reorganize or simplify its neural processing pipeline, as shown in Fig.~\ref{fig:tango_variants}. In particular, the stage-wise results indicate that the intermediate DNN outputs do not fully determine the final enhancement quality: several metrics remain limited, or even decrease, after the neural mask-estimation stage, whereas the largest performance gains are obtained after the spatial filtering stage. This suggests that the downstream filter is able to exploit the spatial structure of the binaural signals and compensate for imperfect mask estimates. Since the first filtering stage in the original TANGO operates only on microphones within the same ear, whose spatial diversity is relatively limited, we investigate whether an earlier multi-node stage and a simplified MN-only configuration can better exploit interaural information while reducing computational cost.

\subsection{End-to-End Training}
\label{ssec:end_to_end_training}

The original TANGO training strategy optimizes the neural mask estimators using mask-level objectives. However, in hybrid neural-beamforming systems, accurate mask estimation does not necessarily translate into optimal enhanced signals after spatial filtering. To better align optimization with the final enhancement objective, we investigate end-to-end training, where the spatial filtering stage is included in the training loop. During training, we use a differentiable implementation of the SDW-MWF, which allows gradients to propagate from the enhanced STFT loss back to the neural mask estimators. At inference time, the spatial filtering stage is computed using the GEVD-based implementation adopted in the original TANGO model. Thus, the reported enhancement results are obtained with the GEVD-based spatial filter, unless explicitly stated otherwise.

%During training, a differentiable SDW-MWF is used as the spatial filter to enable gradient propagation through the beamforming stage. At inference time, the final filtering stage is no longer constrained by differentiability. Unless stated otherwise, inference is therefore performed with GEVD, following TANGO's evaluation setup. %For MN-TANGO, SDW-MWF inference is also reported for comparison.

The end-to-end objective combines the mask-level loss with an enhanced-STFT loss computed after the last spatial filtering stage. Let $\tilde{M}_c$ and $M_c$ denote the estimated and target masks for ear $c\in\{\mathrm{L},\mathrm{R}\}$, and let $\tilde{S}_c$ and $S_c$ denote the corresponding enhanced and clean STFTs. The mask loss is defined as:
\begin{equation}
\mathcal{L}_{\mathrm{mask}}
=
\frac{1}{2}
\sum_{c \in \{\mathrm{L},\mathrm{R}\}}
\mathrm{MSE}\left(\tilde{M}_{c},M_{c}\right).
\label{eq:task_mask_loss}
\end{equation}

%%%%%%%%%%%%%%%%%%%%%%%%%%%%%%%%%%%%%%%%%%%%%%%

\noindent Inspired by~\cite{rong_gtcrn_2024}, the enhanced-STFT loss is defined as:
\begin{equation}
\begin{array}{@{}l@{}}
\ell_{\mathrm{STFT}}\left(\tilde{S}_{c},S_{c}\right)
=
(1-\beta)
\mathrm{MSE}\left(|\tilde{S}_{c}|,|S_{c}|\right)+
\\[0.5ex]
{}
\beta
\left(
\mathrm{MSE}\left(\mathrm{Re}\{\tilde{S}_{c}\},\mathrm{Re}\{S_{c}\}\right)
+
\mathrm{MSE}\left(\mathrm{Im}\{\tilde{S}_{c}\},\mathrm{Im}\{S_{c}\}\right)
\right).
\end{array}
\label{eq:single_ear_stft_loss}
\end{equation}
where $\mathrm{Re}\{\cdot\}$ and $\mathrm{Im}\{\cdot\}$ denote the real and imaginary parts, respectively. Here, \(\beta\) balances magnitude-domain and complex-domain reconstruction terms. The loss is then averaged over the two ears:
\begin{equation}
\mathcal{L}_{\mathrm{STFT}}
=
\frac{1}{2}
\sum_{c \in \{\mathrm{L},\mathrm{R}\}}
\ell_{\mathrm{STFT}}\left(\tilde{S}_{c},S_{c}\right).
\label{eq:task_stft_loss}
\end{equation}

\noindent The full end-to-end objective is:
\begin{equation}
\mathcal{L}_{\mathrm{task}}
=
\alpha \mathcal{L}_{\mathrm{mask}}
+
(1-\alpha)\mathcal{L}_{\mathrm{STFT}},
\label{eq:e2e_loss}
\end{equation}
where $\alpha$ controls the balance between mask reconstruction and final enhanced-signal reconstruction. 

\begin{table*}[t]
    \centering
    
    \caption{Stage-wise performance and complexity comparison of end-to-end trained TANGO variants in FP32.
    Bold and underlined values indicate the best and second-best scores, respectively, among the final GEVD filtering rows.}

    \label{tab:training_variants_metrics}

    \begin{threeparttable}
    \resizebox{\textwidth}{!}{%
    \begin{tabular}{
        cccc
        *{5}{cc}
    }
    \toprule

        \multirow{2}{*}{\textbf{Method}}
    & \multirow{2}{*}{\textbf{MMACs/s}\down}
    & \multirow{2}{*}{\textbf{\#Params}\down}
    & \multirow{2}{*}{\textbf{Step}}
    & \multicolumn{2}{c}{\textbf{SI-SIR}\up}
    & \multicolumn{2}{c}{\textbf{SI-SDR}\up}
    & \multicolumn{2}{c}{\textbf{SI-SAR}\up}
    & \multicolumn{2}{c}{\textbf{STOI}\up}
    & \multicolumn{2}{c}{\textbf{PESQ}\up} \\\cmidrule(lr){5-6}\cmidrule(lr){7-8}
\cmidrule(lr){9-10}\cmidrule(lr){11-12}\cmidrule(lr){13-14}
    
    &
    &
    &
    & \LR
    & \LR
    & \LR
    & \LR
    & \LR \\\midrule

    Noisy
    & $-$
    & $-$
    & $-$
    & $0.0$ & $-4.0$
    & $-0.6$ & $-4.6$
    & $-$ & $-$
    & $0.68$ & $0.56$
    & $1.14$ & $1.10$ \\\midrule

    \multirow{4}{*}{TANGO}
    & \multirow{4}{*}{$65.65$}
    & \multirow{4}{*}{$1~M$}
    & SN-DNN
    & $3.1$ & $0.0$
    & $0.7$ & $-2.2$
    & $7.3$ & $5.5$
    & $0.71$ & $0.59$
    & $1.14$ & $1.09$ \\

    &  &  & Filter$_1$ (GEVD)
    & $9.4$ & $6.7$
    & $-0.7$ & $-2.1$
    & $0.8$ & $0.4$
    & $0.73$ & $0.66$
    & $1.21$ & $1.16$ \\

    &  &  & MN-DNN
    & $13.0$ & $7.8$
    & $5.0$ & $2.2$
    & $6.2$ & $4.6$
    & $0.75$ & $0.74$
    & $1.22$ & $1.15$ \\

    &  &  & Filter$_2$ (GEVD)
    & $\best{24.3}$ & $\best{25.6}$
    & $5.3$ & $\second{4.9}$
    & $5.5$ & $5.0$
    & $0.85$ & $\best{0.85}$
    & $1.76$ & $1.68$ \\\midrule

    \multirow{4}{*}{\shortstack[c]{Inverted TANGO\\($B^\dagger)$}}
    & \multirow{4}{*}{$65.65$}
    & \multirow{4}{*}{$1~M$}
    & MN-DNN
    & $3.3$ & $1.7$
    & $-0.9$ & $-2.2$
    & $4.8$ & $3.5$
    & $0.52$ & $0.58$
    & $1.11$ & $1.09$ \\

    &  &  & Filter$_1$ (GEVD)
    & $12.3$ & $15.2$
    & $-1.3$ & $-0.4$
    & $0.5$ & $0.7$
    & $0.70$ & $0.77$
    & $1.29$ & $1.30$ \\

    &  &  & SN-DNN
    & $9.1$ & $3.9$
    & $3.6$ & $0.3$
    & $6.0$ & $5.1$
    & $0.72$ & $0.67$
    & $1.15$ & $1.10$ \\

    &  &  & Filter$_2$ (GEVD)
    & $\second{24.2}$ & $\second{24.9}$
    & $5.2$ & $\second{4.9}$
    & $5.4$ & $5.1$
    & $0.85$ & $0.84$
    & $1.71$ & $1.67$ \\\midrule

    \multirow{4}{*}{\shortstack[c]{Inverted TANGO\\($B^\star$)}}
    & \multirow{4}{*}{$65.65$}
    & \multirow{4}{*}{$1~M$}
    & MN-DNN
    & $3.1$ & $0.8$
    & $-0.20$ & $-2.1$
    & $6.2$ & $4.7$
    & $0.57$ & $0.55$
    & $1.11$ & $1.10$ \\

    &  &  & Filter$_1$ (GEVD)
    & $11.2$ & $8.7$
    & $4.8$ & $2.3$
    & $6.6$ & $4.4$
    & $0.74$ & $0.62$
    & $1.23$ & $1.13$ \\

    &  &  & SN-DNN
    & $11.2$ & $8.7$
    & $4.8$ & $2.3$
    & $6.6$ & $4.4$
    & $0.74$ & $0.62$
    & $1.25$ & $1.13$ \\

    &  &  & Filter$_2$ (GEVD)
    & $23.2$ & $23.6$
    & $\best{6.7}$ & $\best{5.5}$
    & $\best{7.0}$ & $\best{5.8}$
    & $\best{0.88}$ & $\second{0.84}$
    & $\best{1.84}$ & $\best{1.77}$ \\\midrule

    \multirow{3}{*}{MN-TANGO}
    & \multirow{3}{*}{$\best{30.79}$}
    & \multirow{3}{*}{$\best{0.5~M}$}
    & MN-DNN
    & $12.2$ & $8.9$
    & $4.2$ & $2.0$
    & $5.5$ & $3.8$
    & $0.67$ & $0.61$
    & $1.19$ & $1.13$ \\

    &  &  & Filter$_2$ (GEVD)
    & $23.7$ & $24.2$
    & $\second{6.1}$ & $\best{5.5}$
    & $\second{6.3}$ & $\second{5.6}$
    & $\second{0.86}$ & $\second{0.84}$
    & $\second{1.79}$ & $\second{1.73}$ \\

    &  &  & Filter$_2$ (SDW-MWF)
    & $12.5$ & $10.4$
    & $6.9$ & $5.5$
    & $9.2$ & $8.3$
    & $0.83$ & $0.76$
    & $1.56$ & $1.37$ \\\bottomrule

    \end{tabular}%
    }
    \end{threeparttable}
\end{table*} 

%\subsection{Quantization of MN-TANGO with Knowledge Distillation}
\subsection{Low-Precision TANGO}
\label{ssec:quant_kd}

We evaluate TANGO under low-precision inference using QAT, following Section~\ref{quantized_tango}. Because low-precision inference may degrade the quality of the predicted masks and the resulting enhanced signal, we further introduce knowledge distillation (KD) with floating-point TANGO as the teacher and the quantized model as the student. The KD objective combines mask-level MSE matching and enhanced-STFT matching between the teacher and student outputs. The mask-level KD loss is defined as: 
\begin{equation}
\mathcal{L}_{\mathrm{KD}}^{\mathrm{mask}}
=
\frac{1}{2}
\sum_{c \in \{\mathrm{L},\mathrm{R}\}}
\mathrm{MSE}
\left(
M^{(s)}_{c},
M^{(t)}_{c}
\right),
\label{eq:kd_mask}
\end{equation}
where $M^{(s)}_c$ and $M^{(t)}_c$ denote the final masks predicted by the student and teacher, respectively. The enhanced-STFT KD loss reuses the per-ear STFT loss from Section~\ref{ssec:end_to_end_training}:
\begin{equation}
\mathcal{L}_{\mathrm{KD}}^{\mathrm{STFT}}
=
\frac{1}{2}
\sum_{c \in \{\mathrm{L},\mathrm{R}\}}
\ell_{\mathrm{STFT}}
\left(
Y^{(s)}_{c},Y^{(t)}_{c}
\right),
\label{eq:kd_stft}
\end{equation}
where $Y^{(s)}_c$ and $Y^{(t)}_c$ denote the corresponding enhanced STFTs after the differentiable spatial filter. The final distillation and training objectives are:
\begin{align}
    \mathcal{L}_{\mathrm{distill}}
    &=
    \lambda_{\mathrm{KD}} \mathcal{L}_{\mathrm{KD}}^{\mathrm{mask}}
    +(1-\lambda_{\mathrm{KD}})\mathcal{L}_{\mathrm{KD}}^{\mathrm{STFT}},\\
    \mathcal{L}{\mathrm{total}}
    &=
    \lambda_{\mathrm{task}} \mathcal{L}_{\mathrm{task}}
    +(1-\lambda_{\mathrm{task}})\mathcal{L}_{\mathrm{distill}}.
    \label{eq:kd_total_mask_stft}
\end{align}
where $\lambda_{\mathrm{KD}}$ controls the balance between mask-level and enhanced-STFT distillation, and $\lambda_{\mathrm{task}}$ controls the balance between supervised task training and knowledge distillation.

%\subsection{Low memory, low compute MN-Tango}
\subsection{Low-Compute TANGO}
\label{ssec:efficient_tango_validation}
Beyond low-precision quantization, we further reduce the complexity of TANGO through architectural compression. This allows us to assess whether the proposed quantization pipeline remains effective under stricter compute and memory constraints.

This design adapts the architectural compression strategy introduced in our previous low-latency, low-compute RT-TANGO framework~\cite{benslimane2026rttangorealtimedistributedbinaural}, relying on ERB feature compression to exploit perceptual frequency redundancy and grouped recurrent processing to reduce recurrent computation.

After the point-wise channel-mixing layer, the linear-frequency STFT representation is projected onto a compact ERB scale, reducing the recurrent input dimension. The predicted ERB-domain mask is then mapped back to the original STFT frequency bins before filtering.

The original MN-DNN recurrent block is replaced with grouped LSTM layers. The hidden representation is partitioned into (G) groups processed independently within each recurrent layer. Deterministic interleaving is applied between layers to exchange information across groups~\cite{gao-etal-2018-efficient}.

\section{Experimental setup}
\subsection{Training and Evaluation}
The training data consisted of simulated binaural mixtures generated according to the setup described by Monir et al.~\cite{monir2025frequencyweightedtraininglossesphonemelevel}. The simulated hearing-aid configuration contains four microphones, two placed on each ear. Speech signals were taken from LibriSpeech~\cite{7178964} and combined with speech-shaped noise and real environmental noise sources. For evaluation, we used a subset of the BinauRec dataset\footnote{Online: https://zenodo.org/records/7256984}~\cite{delebecque_binaurec_2023}. This subset contains 1,200 binaural mixtures generated from measured room impulse responses. The RIRs were recorded using a portable hearing laboratory equipped with behind-the-ear hearing-aid shells mounted on a dummy head~\cite{pavlovic_high-fidelity_2019}. Mixtures were generated at input SNRs of $-5$, $0$, and $5$~dB. The target source was placed in front of the listener, while the noise source was located either $45^\circ$ or $90^\circ$ to the right of the target. This setup reflects a typical hearing-aid scenario, with a frontal target speaker and lateral interfering noise. Only right-side noise locations are considered, since left-side configurations are symmetric, making the right ear more challenging. 

\begin{table*}[t]
\centering

\caption{Effect of W8A8 quantization and knowledge distillation on MN-TANGO before and after GEVD filtering.}
\label{tab:tango_qat_bf_kd_all_metrics}

\begin{threeparttable}

\begin{tabular}{
lcc
*{5}{cc} % Metrics (SI-SIR, ...)
}
\toprule

\multirow{2}{*}{\textbf{Output}}
& \multirow{2}{*}{\textbf{Precision}}
& \multirow{2}{*}{\textbf{KD}}
& \multicolumn{2}{c}{\textbf{SI-SIR}\up}
& \multicolumn{2}{c}{\textbf{SI-SDR}\up}
& \multicolumn{2}{c}{\textbf{SI-SAR}\up}
& \multicolumn{2}{c}{\textbf{STOI}\up}
& \multicolumn{2}{c}{\textbf{PESQ}\up} \\
\cmidrule(lr){4-5}
\cmidrule(lr){6-7}
\cmidrule(lr){8-9}
\cmidrule(lr){10-11}
\cmidrule(lr){12-13}

&
&
& \LR
& \LR
& \LR
& \LR
& \LR \\
\midrule

\multirow{3}{*}{MN-DNN}
& FP32
& $-$
& $12.2$ & $8.9$
& $4.2$ & $2.0$
& $5.5$ & $3.8$
& $0.67$ & $0.61$
& $1.19$ & $1.13$ \\

& W8A8
& \xmark
& $10.7$ & $7.1$
& $3.7$ & $1.4$
& $5.4$ & $4.0$
& $0.66$ & $0.59$
& $1.18$ & $1.12$ \\

& W8A8
& \cmark
& $10.6$ & $7.0$
& $3.6$ & $1.3$
& $5.3$ & $3.9$
& $0.65$ & $0.59$
& $1.18$ & $1.12$ \\

\midrule

\multirow{3}{*}{\shortstack[c]{Final output\\(GEVD)}}
& FP32
& $-$
& $23.7$ & $24.2$
& $6.1$ & $5.5$
& $6.3$ & $5.6$
& $0.86$ & $0.84$
& $1.79$ & $1.73$ \\

& W8A8
& \xmark
& $23.6$ & $24.8$
& $5.8$ & $5.4$
& $6.1$ & $5.5$
& $0.86$ & $0.84$
& $1.77$ & $1.71$ \\

& W8A8
& \cmark
& $23.9$ & $25.2$
& $5.8$ & $5.3$
& $6.0$ & $5.5$
& $0.86$ & $0.84$
& $1.77$ & $1.72$ \\

\bottomrule

\end{tabular}

%\begin{tablenotes}[flushleft]
%    \footnotesize
%    \item FP32 denotes the non-quantized model. QAT denotes the quantization-aware trained %model. KD denotes knowledge distillation. $KD$ denotes weight-level distillation, while %$KD^{\mathrm{SI\mbox{-}SDR}}$ denotes SI-SDR-based distillation.
%\end{tablenotes}

\end{threeparttable}
\end{table*}

\subsection{Model Configuration}
\label{sssec:model_configuration}

All mask estimators use $257$-bin magnitude STFT features computed with a $512$-point FFT, a $512$-sample Hann window, and a $256$-sample hop at $16$~kHz, corresponding to $62.5$ frames/s. Floating-point end-to-end models are trained on $512$-frame segments, while grouped QAT models are initialized from their floating-point grouped checkpoints and fine-tuned on $64$-frame segments, to reduce training time.

The learning rates are $5\times10^{-4}$ for floating-point training and $10^{-4}$ for QAT fine-tuning. For the end-to-end objective, we set $\alpha=0.3$ and $\beta=0.3$. For the KD experiments, we set $\lambda_{\mathrm{KD}}=0.3$ and $\lambda_{\mathrm{task}}=0.7$. In all experiments, all spatial filtering stages use the trade-off parameter \(\mu=1\).

TANGO uses two recurrent mask estimators. The SN-DNN is a unidirectional LSTM mask estimator with three LSTM layers of $128$ hidden units and a fully connected mask head with hidden/output dimensions $256$ and $257$. The MN-DNN first applies a point-wise convolution to combine the two input channels, followed by GELU activation, layer normalization, a three-layer unidirectional LSTM with $128$ hidden units, and a fully connected sigmoid mask head with $257$ output bins.

For the low-compute TANGO, the ERB projection uses \(64\) low-frequency linear bins and \(64\) ERB bands, yielding a \(128\)-dimensional recurrent input for most grouped models. When divisibility by the group count \(G\) is required, the number of ERB bands is adjusted accordingly. The grouped recurrent block consists of two unidirectional LSTM layers with \(128\) hidden units, and the final mask is mapped back to the original \(257\) STFT frequency bins before filtering.

\subsection{Quantization Configurations}

In this work, quantization is applied only to the neural mask estimators. All fixed signal-processing components, including STFT/iSTFT operations, covariance estimation, and SDW-MWF/GEVD spatial filtering, remain in floating-point precision (FP32). Although the low-precision methodology can be applied to any TANGO variant, the QAT and KD experiments in this paper are conducted on MN-TANGO. 
DPTQ is implemented using the PyTorch eager-mode dynamic quantization API.\footnote{\texttt{torch.ao.quantization.quantize\_dynamic}.} For QAT, we use a custom implementation inspired by the FQSS fully quantized source-separation framework\footnote{Code: \url{https://github.com/ssi-research/FQSS/tree/main}.}. Weights use a symmetric signed quantizer, while activations are quantized using an asymmetric affine quantizer whose range is initialized from observed activation minima and maxima. Observer-based range updates are enabled during an initial warm-up phase and then frozen, after which quantization ranges are optimized by gradient descent. The main QAT configuration uses mixed precision: trainable weights and internal activations are quantized to $8$ bits (W8A8), while input and output mask tensors are simulated using $16$-bit. 
Bias terms are kept in higher precision and added in the accumulator domain before requantization. Throughout this work, W8A8 refers specifically to the precision of internal neural layers rather than to input/output mask tensors.

\subsection{Evaluation Metrics}
Enhancement performance was measured using scale-invariant signal-to-distortion ratio (SI-SDR), scale-invariant signal-to-interference ratio (SI-SIR), and scale-invariant signal-to-artifacts ratio (SI-SAR), all reported in dB~\cite{8683855}, as well as the short-time objective intelligibility (STOI)~\cite{5495701} and perceptual evaluation of speech quality (PESQ)~\cite{941023}. The unprocessed noisy input mixture is reported in the result tables to provide a reference signal quality prior to enhancement. SI-SAR is not reported for this baseline, as the metric reflects artifacts introduced by signal processing and is therefore not applicable to an unprocessed input.
Computational complexity and model size are reported per processing node, with complexity measured in multiply-accumulate operations (MACs) and model size given in terms of the number of trainable parameters and memory footprint.

\section{Results}
%%%%%%%%%%%%%%%%%%%%%%%%%%%%%%%%%%%%%%%%%%%%%%%%%%%%%%%%%%%%%%%%%%%%%%%%%%%%%%%%%%%%%%%%%%%%%
%\textcolor{red}{Martyna: Il me manque ici main findings, intrepretation, pour le moment c'est tres table-driven.}
%%%%%%%%%%%%%%%%%%%%%%%%%%%%%%%%%%%%%%%%%%%%%%%%%%%%%%%%%%%%%%%%%%%%%%%%%%%%%%%%%%%%%%%%%%%%%

%\subsection{Quantization of baseline TaANGO}
\subsection{Quantized TANGO}

Table~\ref{tab:quant_lr_metrics} compares DPTQ and QAT without KD on the full TANGO model, revealing a clear performance gap between post-training and training-aware quantization. DPTQ leads to a noticeable degradation compared with the FP32 baseline. This likely reflects the quantization sensitivity of LSTM layers, whose activations and internal states can span different dynamic ranges. %and in frame-wise streaming inference, where dynamic activation ranges may be estimated from small tensors.
In contrast, weight-only QAT preserves the FP32 performance almost exactly, indicating that TANGO can effectively adapt to quantization noise when low-precision effects are incorporated during training.
When activations and I/O tensors are also quantized, performance slightly decreases, mainly in SI-SDR and SI-SAR, while STOI and PESQ remain close to the FP32 baseline. This suggests that the neural mask estimators are highly sensitive to naïve post-training quantization, but can effectively adapt to low-precision constraints when quantization noise is incorporated during training.

%%%%%%%%%%%%%%%%%%%%%%%%%%%%%%%%%%%%%%%%%%%%%%%%%%%%%%%%%%%%%%%%%%%%%%%%%%%%%%%%%%%%%%%%%%%%%%%% !!!!!!!!!!!!!!!!!!!!!!!
%\textcolor{red}{Manque: Pourquoi DPTQ échoue sur TANGO ?: probablement car LSTM + mask estimation sensibles aux erreurs de poids, Pourquoi QAT récupère ?: adaptation des masques pendant entraînement. } \zahra{done, check blue}

%%%%%%%%%%%%%%%%%%%%%%%%%%%%%%%%%%%%%%%%%%%%%%%%%%%%%%%%%%%%%%%%%%%%%%%%%%%%%%%%%%%%%%%%%%%%%%%%

\comment{
\begin{figure}[t]
  \centering
  \includegraphics[width=\columnwidth]{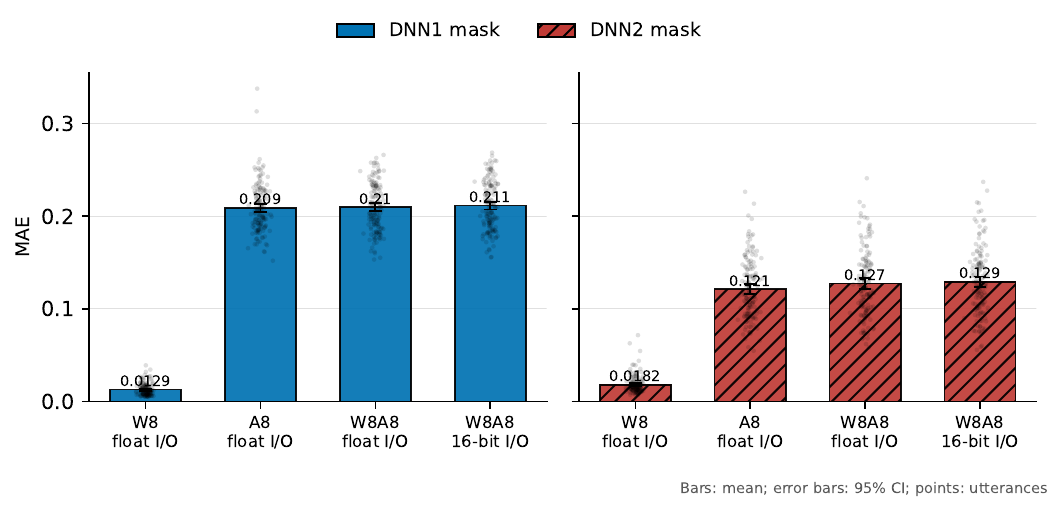}
  \caption{My PDF graphic}
  \label{fig:pdfgraphic}
\end{figure}
}

\subsection{Quantized MN-TANGO}
\label{ssec:quantized_MN_tango}

%%%%%%%%%%%%%%%%%%%%%%%%%%%%%%%%%%%%%%%%%%%%%%%%%%%%%%%%%%%%%%%%%%%%%%%%%%%%%%%%%%%%%%%%%%%%%%%%
% REMARQUE MARTYNA
%\textcolor{red}{Beaucoup trop de texte, très dense. Mauvais ordre narratif. Trois résultats majeurs sont entremêlés. A vrai dire, il y a 3 mains claims ici: 1) Le spatial filter final apporte l’essentiel du gain. 2) MN-TANGO est le meilleur tradeoff. 3) Le GEVD filter absorbe largement les erreurs de quantization. Enfin Partie B-star a reduire, Partie MN-Tango a renforcer} \zahra{agree}\\
%%%%%%%%%%%%%%%%%%%%%%%%%%%%%%%%%%%%%%%%%%%%%%%%%%%%%%%%%%%%%%%%%%%%%%%%%%%%%%%%%%%%%%%%%%%%%%%%

%\martyna{Three main observations emerge from Tables~\ref{tab:training_variants_metrics}
%and~\ref{tab:tango_qat_bf_kd_all_metrics}: (i) the final spatial filtering stage provides most of the enhancement gains, (ii) MN-TANGO offers the best complexity-performance trade-off, and (iii) GEVD filtering compensates for most quantization-induced degradation.}

%%%%%%%%%%%%%%%%%%%%%%%%%%%%%%%%%%%%%%%%%%%%%%%%%%%%%%%%%%%%%%%%%%%%%%%%%%%%%%%%%%%%%%%%%
%\zahra{très bien résumé, tu voudra mettre ça just en dessous de V. RESULTS ? , commemain finding ? ou dans l'intro maybe ?} \textcolor{red}{Plus haut dans le resultats pas trop car cela fait la reference a des tableaux cités apres la sousection A. Mais je pense que cela peut etre deplacer dans l'intro (sans ref tableaux) comme main contributions! et aussi mentionner dans l'abstract en version tres short}
%%%%%%%%%%%%%%%%%%%%%%%%%%%%%%%%%%%%%%%%%%%%%%%%%%%%%%%%%%%%%%%%%%%%%%%%%%%%%%%%%%%%%%%%%

\begin{table}[t]
    \centering
    \caption{Component-wise computational complexity of the grouped recurrent architecture. Component costs and frame totals are in kMACs/frame; the last column is in MMAC/s.}
    \label{tab:group_complexity}
    \resizebox{\columnwidth}{!}{%
    \begin{tabular}{ccccccc}\toprule
        \textbf{G} & \textbf{PW} & \textbf{LSTM} & \textbf{ERB+Inv.} & \textbf{FC} & \textbf{Total/frame}\down & \textbf{Total/s}\down \\\midrule
        $1$  & $0.51$ & $459.26$ & $-$     & $32.90$ & $492.67$ & $30.79$ \\
        $2$  & $0.51$ & $131.07$ & $24.70$ & $16.38$ & $172.67$ & $10.79$ \\
        $4$  & $0.51$ & $65.54$  & $24.70$ & $16.38$ & $107.14$ & $6.70$  \\
        $6$  & $0.51$ & $42.34$  & $23.93$ & $15.88$ & $82.66$  & $5.17$  \\
        $8$  & $0.51$ & $32.77$  & $24.70$ & $16.38$ & $74.37$  & $4.65$  \\
        $10$ & $0.51$ & $27.04$  & $25.48$ & $16.90$ & $69.93$  & $4.37$  \\\bottomrule
    \end{tabular}%
    }
\end{table}

%%%%%%%%%%%%%%%%%%%%%%%%%%%%%%%%%%%%%%%%%%%%%%%%%%%%%%%%%%%%%%%%%%%%%%%
%%%%%%%%%%%%%%%%%%%%%%%%%%%%%%%%%%%%%%%%%%%%%%%%%%%%%%%%%%%%%%%%%%%%%%%
\comment{

\begin{table}[H]
\centering
\caption{Component-wise computational complexity of the grouped recurrent architecture.}
\label{tab:group_mmacs}
\small
\setlength{\tabcolsep}{4pt}
\begin{tabular}{c c c c c c}
\hline
\textbf{Groups} & \textbf{PW} & \textbf{LSTM} & \textbf{ERB + Inv. ERB} & \textbf{FC} & \textbf{Total} \\
\hline
1  & 0.032 & 28.704 & 0.000 & 2.056 & 30.792 \\
2  & 0.032 & 8.192  & 1.544 & 1.024 & 10.792 \\
4  & 0.032 & 4.096  & 1.544 & 1.024 & 6.696  \\
6  & 0.032 & 2.646  & 1.496 & 0.992 & 5.166  \\
8  & 0.032 & 2.048  & 1.544 & 1.024 & 4.648  \\
10 & 0.032 & 1.690  & 1.592 & 1.056 & 4.371  \\
\hline
\end{tabular}
\end{table}

%%%%%%%%%%%%%%%%%%%%%%%%%%%%%%%%%%%%%%%%%%%%%%%%%%%%%%%%%%

\begin{table}[H]
\centering
\caption{Component-wise kMACs/frame of the grouped recurrent architecture.}
\label{tab:group_kmacs_frame}
\small
\setlength{\tabcolsep}{4pt}
\begin{tabular}{c c c c c c}
\hline
\textbf{Groups} & \textbf{PW} & \textbf{LSTM} & \textbf{ERB + Inv. ERB} & \textbf{FC} & \textbf{Total} \\
\hline
1  & 0.514 & 459.264 & 0.000  & 32.896 & 492.674 \\
2  & 0.514 & 131.072 & 24.704 & 16.384 & 172.674 \\
4  & 0.514 & 65.536  & 24.704 & 16.384 & 107.138 \\
6  & 0.514 & 42.336  & 23.932 & 15.876 & 82.658  \\
8  & 0.514 & 32.768  & 24.704 & 16.384 & 74.370  \\
10 & 0.514 & 27.040  & 25.476 & 16.900 & 69.930  \\
\hline
\end{tabular}
\end{table}

\begin{table}[H]
\centering
\caption{Component-wise computational complexity of the grouped recurrent architecture.}
\label{tab:group_complexity}
\small
\setlength{\tabcolsep}{4pt}
\begin{tabular}{c c c c c c c}
\hline
\textbf{G} 
& \textbf{PW} 
& \textbf{LSTM} 
& \textbf{ERB+Inv.} 
& \textbf{FC} 
& \textbf{Total/frame}
& \textbf{Total/s} \\
\hline
1  & 0.514 & 459.264 & 0.000  & 32.896 & 492.674 & 30.792 \\
2  & 0.514 & 131.072 & 24.704 & 16.384 & 172.674 & 10.792 \\
4  & 0.514 & 65.536  & 24.704 & 16.384 & 107.138 & 6.696  \\
6  & 0.514 & 42.336  & 23.932 & 15.876 & 82.658  & 5.166  \\
8  & 0.514 & 32.768  & 24.704 & 16.384 & 74.370  & 4.648  \\
10 & 0.514 & 27.040  & 25.476 & 16.900 & 69.930  & 4.371  \\
\hline
\end{tabular}
\end{table}

\begin{table}[H]
\centering
\caption{Component-wise MAC count per frame of the MN-DNN architecture for different numbers of groups. The LSTM column reports the sum over all LSTM layers.}
\label{tab:group_params}
\begin{tabular}{c c c c c c c}
\hline
\textbf{Groups} & \textbf{PW} & \textbf{LSTM} & \textbf{ERB} & \textbf{Inv. ERB} & \textbf{FC} & \textbf{Total} \\
\hline
1  & 514 & 459,264 & 0      & 0      & 32,896 & 492,674 \\
2  & 514 & 131,072 & 12,352 & 12,352 & 16,384 & 172,674 \\
4  & 514 & 65,536  & 12,352 & 12,352 & 16,384 & 107,138 \\
6  & 514 & 42,336  & 11,966 & 11,966 & 15,876 & 82,658  \\
8  & 514 & 32,768  & 12,352 & 12,352 & 16,384 & 74,370  \\
10 & 514 & 27,040  & 12,738 & 12,738 & 16,900 & 69,930  \\
\hline
\end{tabular}
\end{table}

\begin{table}[H]
\centering
\caption{Computational complexity of the grouped recurrent architecture for different numbers of groups.}
\label{tab:group_mmacs}
\begin{tabular}{c c c c c c c}
\hline
\textbf{Groups} & \textbf{PW} & \textbf{LSTM} & \textbf{ERB} & \textbf{Inv. ERB} & \textbf{FC} & \textbf{Total} \\
\hline
1  & 0.032125 & 28.704 & 0.000000 & 0.000000 & 2.05600 & 30.792125 \\
2  & 0.032125 & 8.192  & 0.772000 & 0.772000 & 1.02400 & 10.792125 \\
4  & 0.032125 & 4.096  & 0.772000 & 0.772000 & 1.02400 & 6.696125  \\
6  & 0.032125 & 2.646  & 0.747875 & 0.747875 & 0.99225 & 5.166125  \\
8  & 0.032125 & 2.048  & 0.772000 & 0.772000 & 1.02400 & 4.648125  \\
10 & 0.032125 & 1.690  & 0.796125 & 0.796125 & 1.05625 & 4.370625  \\
\hline
\end{tabular}
\end{table}

\begin{table}[t]
\centering
\caption{Computational complexity for different numbers of groups.}
\label{tab:group_macs}
\begin{tabular}{c c c}
\hline
\textbf{Groups} & \textbf{LSTM MACs} & \textbf{Total MACs} \\
\hline
1  & -- & -- \\
2  & -- & -- \\
4  & -- & -- \\
6  & -- & -- \\
8  & -- & -- \\
10 & -- & -- \\
\hline
\end{tabular}
\end{table}

}
\begin{table*}[t]
    \centering
    \caption{Performance-complexity trade-off of W8A8 MN-TANGO variants with grouped recurrent layers.}
    \label{tab:w8a8_mntango_grouping_tradeoff}

    \begin{threeparttable}
    
    \resizebox{\textwidth}{!}{%
    \begin{tabular}{
        cccccc
        *{5}{cc}
    }\toprule
    
    \multirow{2}{*}{\textbf{Method}}
    & \multirow{2}{*}{\textbf{G}}
    & \multirow{2}{*}{\textbf{MMACs/s}\down}
    & \multirow{2}{*}{\textbf{\#Params}\down}
    & \multirow{2}{*}{\textbf{Memory}\down}
    & \multirow{2}{*}{\textbf{Step}}
    & \multicolumn{2}{c}{\textbf{SI-SIR}\up}
    & \multicolumn{2}{c}{\textbf{SI-SDR}\up}
    & \multicolumn{2}{c}{\textbf{SI-SAR}\up}
    & \multicolumn{2}{c}{\textbf{STOI}\up}
    & \multicolumn{2}{c}{\textbf{PESQ}\up} \\
    \cmidrule(lr){7-8}
    \cmidrule(lr){9-10}
    \cmidrule(lr){11-12}
    \cmidrule(lr){13-14}
    \cmidrule(lr){15-16}
    
    &
    &
    &
    &
    &
    & \LR
    & \LR
    & \LR
    & \LR
    & \LR \\\midrule

    Noisy
    & $-$
    & $-$
    & $-$
    & $-$ 
    & $-$
    & $0.0$ & $-4.0$
    & $-0.6$ & $-4.6$
    & $-$ & $-$
    & $0.68$ & $0.56$
    & $1.14$ & $1.10$ \\\midrule
    
    TANGO
    & $-$
    & $65.65$
    & $1M$
    & $4.03~MB$ 
    & Filter$_2$ (GEVD)
    & $\best{24.3}$ & $\best{25.6}$
    & $5.3$ & $4.9$
    & $5.5$ & $5.0$
    & $\second{0.85}$ & $\best{0.85}$
    & $\second{1.76}$ & $\second{1.68}$ \\\midrule

    \multirow{2}{*}{\shortstack[c]{MN-TANGO\\W8A8}}
    & \multirow{2}{*}{$-$}
    & \multirow{2}{*}{$30.79$}
    & \multirow{2}{*}{$0.5$~M}
    & \multirow{2}{*}{$0.508$~MB}
    & MN-DNN
    & $12.2$ & $8.9$
    & $4.2$ & $2.0$
    & $5.5$ & $3.8$
    & $0.67$ & $0.61$
    & $1.19$ & $1.13$ \\

    &  &  &  &  & Filter$_2$ (GEVD)
    & $\second{23.7}$ & $\second{24.2}$
    & $\best{6.1}$ & $\best{5.5}$
    & $\best{6.3}$ & $\best{5.6}$
    & $\best{0.86}$ & $\second{0.84}$
    & $\best{1.79}$ & $\best{1.73}$ \\\midrule

    \multirow{2}{*}{\shortstack[c]{MN-TANGO\\W8A8}}
    & \multirow{2}{*}{$2$}
    & \multirow{2}{*}{$\second{10.79}$}
    & \multirow{2}{*}{$\second{0.179~M}$}
    & \multirow{2}{*}{$\second{0.274~MB}$}
    & MN-DNN
    & $10.3$ & $6.4$
    & $3.4$ & $1.0$
    & $5.3$ & $3.8$
    & $0.65$ & $0.58$
    & $1.18$ & $1.11$ \\

    &  &  &  &  & Filter$_2$ (GEVD)
    & $22.7$ & $22.8$
    & $\second{5.7}$ & $\second{5.0}$
    & $\second{6.0}$ & $\second{5.3}$
    & $\second{0.85}$ & $0.83$
    & $1.74$ & $1.66$ \\\midrule
    
    \multirow{2}{*}{\shortstack[c]{MN-TANGO\\W8A8}}
    & \multirow{2}{*}{$8$}
    & \multirow{2}{*}{$\best{4.65}$}
    & \multirow{2}{*}{$\best{0.081~M}$}
    & \multirow{2}{*}{$\best{0.177~MB}$}
    & MN-DNN
    & $9.1$ & $5.7$
    & $2.7$ & $0.4$
    & $4.9$ & $3.6$
    & $0.62$ & $0.55$
    & $1.16$ & $1.10$ \\

    &  &  &  &  & Filter$_2$ (GEVD)
    & $21.2$ & $21.3$
    & $5.2$ & $4.4$
    & $5.6$ & $4.8$
    & $0.84$ & $0.82$
    & $1.68$ & $1.60$ \\\bottomrule
    \end{tabular}%
    }
    
    \begin{tablenotes}
        \footnotesize
        \item[*] TANGO is the full-precision reference; grouping and W8A8 quantization are applied only to MN-TANGO variants.
    \end{tablenotes}
    \end{threeparttable}
\end{table*}

\comment{\begin{table*}[t]
    \centering
    \caption{Performance--complexity trade-off of W8A8 MN-Tango variants with grouped recurrent layers.}
    \label{tab:w8a8_mntango_grouping_tradeoff}

    \begin{threeparttable}
    
    \resizebox{\textwidth}{!}{%
    \begin{tabular}{
        cccccc
        *{5}{cc}
    }\toprule
    
    \multirow{2}{*}{\textbf{Method}}
    & \multirow{2}{*}{\textbf{G}}
    & \multirow{2}{*}{\textbf{MMACs/s}\down}
    & \multirow{2}{*}{\textbf{\#Params}\down}
    & \multirow{2}{*}{\textbf{Memory}\down}
    & \multirow{2}{*}{\textbf{Step}}
    & \multicolumn{2}{c}{\textbf{SI-SIR}\up}
    & \multicolumn{2}{c}{\textbf{SI-SDR}\up}
    & \multicolumn{2}{c}{\textbf{SI-SAR}\up}
    & \multicolumn{2}{c}{\textbf{STOI}\up}
    & \multicolumn{2}{c}{\textbf{PESQ}\up} \\
    \cmidrule(lr){7-8}
    \cmidrule(lr){9-10}
    \cmidrule(lr){11-12}
    \cmidrule(lr){13-14}
    \cmidrule(lr){15-16}
    
    &
    &
    &
    &
    &
    & \LR
    & \LR
    & \LR
    & \LR
    & \LR \\\midrule

    Noisy
    & $-$
    & $-$
    & $-$
    & $-$ 
    & $-$
    & $0.0$ & $-4.0$
    & $-0.6$ & $-4.6$
    & $-$ & $-$
    & $0.68$ & $0.56$
    & $1.14$ & $1.10$ \\\midrule
    
    TANGO
    & $-$
    & $65.65$
    & $1M$
    & $4.03$ 
    & Filter$_2$ (GEVD)
    & $24.3$ & $25.6$
    & $5.3$ & $4.9$
    & $5.5$ & $5.0$
    & $0.85$ & $0.85$
    & $1.76$ & $1.68$ \\\midrule

    \multirow{2}{*}{\shortstack[c]{MN-TANGO\\W8A8}}
    & \multirow{2}{*}{$-$}
    & \multirow{2}{*}{$30.79$}
    & \multirow{2}{*}{$0.5$~M}
    & \multirow{2}{*}{$0.508$~MB}
    & MN-DNN
    & $12.2$ & $8.9$
    & $4.2$ & $2.0$
    & $5.5$ & $3.8$
    & $0.67$ & $0.61$
    & $1.19$ & $1.13$ \\

    &  &  &  &  & Filter$_2$ (GEVD)
    & $\best{23.7}$ & $\best{24.2}$
    & $\best{6.1}$ & $\best{5.5}$
    & $\best{6.3}$ & $\best{5.6}$
    & $\best{0.86}$ & $\best{0.84}$
    & $\best{1.79}$ & $\best{1.73}$ \\\midrule

    \multirow{2}{*}{\shortstack[c]{MN-TANGO\\W8A8}}
    & \multirow{2}{*}{$2$}
    & \multirow{2}{*}{$\second{10.79}$}
    & \multirow{2}{*}{$\second{0.179~M}$}
    & \multirow{2}{*}{$\second{0.274~MB}$}
    & MN-DNN
    & $10.3$ & $6.4$
    & $3.4$ & $1.0$
    & $5.3$ & $3.8$
    & $0.65$ & $0.58$
    & $1.18$ & $1.11$ \\

    &  &  &  &  & Filter$_2$ (GEVD)
    & $\second{22.7}$ & $\second{22.8}$
    & $\second{5.7}$ & $\second{5.0}$
    & $\second{6.0}$ & $\second{5.3}$
    & $\second{0.85}$ & $\second{0.83}$
    & $\second{1.74}$ & $\second{1.66}$ \\\midrule
    
    \multirow{2}{*}{\shortstack[c]{MN-TANGO\\W8A8}}
    & \multirow{2}{*}{$8$}
    & \multirow{2}{*}{$\best{4.65}$}
    & \multirow{2}{*}{$\best{0.081~M}$}
    & \multirow{2}{*}{$\best{0.177~MB}$ }
    & MN-DNN
    & $9.1$ & $5.7$
    & $2.7$ & $0.4$
    & $4.9$ & $3.6$
    & $0.62$ & $0.55$
    & $1.16$ & $1.10$ \\

    &  &  &  &  & Filter$_2$ (GEVD)
    & $21.2$ & $21.3$
    & $5.2$ & $4.4$
    & $5.6$ & $4.8$
    & $0.84$ & $0.82$
    & $1.68$ & $1.60$ \\\bottomrule
    \end{tabular}%
    }
    \end{threeparttable}
    
\end{table*}
}

Table~\ref{tab:training_variants_metrics} compares the TANGO variants introduced in Section~\ref{ssec:reduced_e2e}. Across  variants, the largest performance jump consistently occurs after the final spatial filtering stage rather than within the neural mask estimators. This indicates that the downstream spatial filter contributes most of the final enhancement by exploiting binaural spatial structure and compensating for imperfect mask estimates. For the original TANGO model, the final stage increases SI-SIR from $13.0/7.8$~dB after MN-DNN to $24.3/25.6$~dB after GEVD filtering for the left and right ears, respectively. Among the inverted variants, \(B^\dagger\) achieves performance close to the original TANGO model but does not provide consistent improvements across metrics. By contrast, \(B^\star\) achieves the best reconstruction quality in terms of SI-SDR, SI-SAR, STOI, and PESQ, although its SI-SIR remains slightly below that of the original TANGO model. This suggests that performing multi-node processing earlier in the pipeline improves signal reconstruction quality, even if interference suppression is not maximized.

MN-TANGO provides the best overall trade-off between enhancement quality, computational cost, and communication overhead. With GEVD inference, it reaches SI-SIR values of $23.7$~dB and $24.2$~dB for the left and right ears, respectively. Compared with the full TANGO system, MN-TANGO reduces the parameter count and neural computational cost by approximately 50\%, from \(1.0\)M to \(0.5\)M parameters and from \(65.65\) to \(30.79\)~MMAC/s, while preserving the single inter-node exchange required by the original TANGO architecture. 

Since the end-to-end training uses a differentiable SDW-MWF implementation, we additionally evaluate MN-TANGO using the same SDW-MWF implementation at inference. This allows us to quantify the effect of the train–test filtering mismatch and assess whether using the same filtering formulation during training and inference improves performance. The results show that SDW-MWF inference produces higher SI-SDR and SI-SAR, whereas GEVD-based filtering provides stronger interference suppression and better perceptual scores. This suggests that the differentiable SDW-MWF is effective as an optimization surrogate during training, while the GEVD-based implementation remains preferable.

Table~\ref{tab:tango_qat_bf_kd_all_metrics} evaluates the impact of W8A8 quantization and knowledge distillation on MN-TANGO. Although quantization noticeably degrades the intermediate MN-DNN outputs, most of this degradation disappears after GEVD filtering. This indicates that the spatial filtering stage is robust to errors introduced by 8-bit weight and activation quantization. KD provides only marginal improvements, suggesting that teacher guidance brings limited benefit once the downstream spatial filter compensates for most quantization artifacts. %\zahra{romain : Though KD would allows to do QAT on datasets without ground truth (clean/noisy speech pairs). This open quite a few application possibilities and should probably be mentioned here.}

\subsection{Low memory, low compute MN-TANGO }

\begin{figure}[!ht]
  \centering
  \includegraphics[width=\columnwidth]{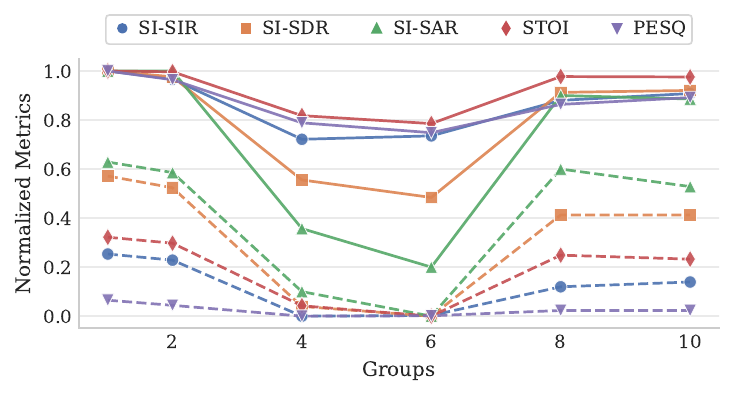}
  \caption{Effect of the number of groups on normalized metrics. Dashed lines correspond to the DNN output, while solid lines represent the final GEVD output. The metrics are averaged over the left and right ears and min-max normalized for compact visualization}
  \label{fig:group_effect}
\end{figure}

Fig.~\ref{fig:group_effect} and Table~\ref{tab:group_complexity} show that recurrent grouping provides substantial computational savings with controlled performance degradation. 
The best performance is obtained with one or two groups, whereas four and six groups lead to a noticeable degradation. Performance partially recovers for eight and ten groups, indicating that the effect of grouping is not strictly monotonic and depends on how the recurrent representation is partitioned. At the same time, increasing the number of groups substantially reduces computational cost: the total complexity decreases from $492.67$~kMACs/frame ($30.79$~MMAC/s) with one group to $69.93$~kMACs/frame ($4.37$~MMAC/s) with ten groups. This reduction is largely driven by the recurrent block, whose complexity decreases from $459.26$ to $27.04$~kMACs/frame, confirming that the LSTM dominates the neural computation. These results indicate that grouped recurrent layers offer large computational savings, but the number of groups must be selected as a trade-off between enhancement quality and efficiency.

Table~\ref{tab:w8a8_mntango_grouping_tradeoff} then evaluates grouped MN-TANGO configurations after W8A8 quantization. The same trend is observed after quantization: increasing the number of recurrent groups reduces both the parameter count and memory footprint, from 0.5~M parameters and 0.508~MB to 0.179~M parameters and 0.274~MB for $G=2$, and to 0.081~M parameters and 0.177~MB for $G=8$. This reduction comes with lower MN-DNN performance, but the GEVD stage consistently improves the final output for all configurations. Among the quantized variants, $G=2$ offers the best complexity-performance trade-off, whereas $G=8$ provides the most compact model with the lowest computational cost. 

\section{Conclusion}
We presented a low-compute quantized version of TANGO for distributed binaural SE. Our study showed that hybrid neural-spatial enhancement systems are particularly well suited to low-precision inference: while quantization degrades neural mask estimation, the spatial filtering stage effectively mitigates most of the resulting errors. Based on this insight, we simplified the original two-stage architecture into MN-TANGO and combined W8A8 quantization, grouped recurrent processing, and ERB compression to significantly reduce memory footprint and computational complexity while maintaining strong enhancement performance. The best trade-off was obtained with two recurrent groups, reducing the complexity from 65.65 to 10.79~MMAC/s, with 0.179M parameters and 0.274~MB memory. The most compact configuration, with eight groups, reduced the complexity further to 4.65~MMAC/s, while using only 0.081M parameters and 0.177~MB memory. These results show that quantized grouped MN-TANGO is a promising architecture for resource-constrained binaural speech enhancement.

%We presented a low-compute quantized version of TANGO for distributed binaural speech enhancement. Our study showed that hybrid neural-spatial enhancement systems are particularly well suited to low-precision inference: while quantization degrades neural mask estimation, the spatial filtering stage effectively mitigates most of the resulting errors. Based on this insight, we simplified the original two-stage architecture into MN-TANGO and combined W8A8 quantization, grouped recurrent processing, and ERB compression to significantly reduce memory footprint and computational complexity while maintaining strong enhancement performance. The best trade-off was obtained with two recurrent groups, reducing complexity to 10.79~MMAC/s\zahra{remind the original is 65MMACs/s} with 0.179M parameters and 0.274 MB~memory. The most compact configuration, with eight groups, further reduced complexity to 4.65 MMAC/s with only 0.081M parameters and 0.177 MB memory, with only moderate performance degradation. These results show that quantized grouped MN-TANGO is a promising architecture for resource-constrained binaural speech enhancement.

%\textcolor{red}{Attention ! Derniere page à ne pas remplir complètement, faut prévoir de la place pour les auteurs si accepted.}

\section*{Acknowledgment}
This research was carried out with the support of the French National Research Agency as part of the REFINED project, ``REal-time artiFicial INtelligence for hEaring aiDs'' (ANR21-CE19-0043).

\section{Generative AI Use Disclosure}
The authors used AI tools solely for language editing and clarity improvement. All scientific ideas, content, and results were developed and verified by the authors.

%\bibliographystyle{IEEEtran}
%\bibliography{references}

%for ArXiv (a mettre à jour si .bib change!!!)

\end{document}